\documentclass[twocolumn,trackchanges]{aastex631}

\usepackage{amsmath,xcolor}
\usepackage{amsmath} 
\usepackage{array}

\begin{document}

\title{Irradiated Atmospheres I: Heating by Vertical-Mixing Induced Energy Transport}

\correspondingauthor{Cong Yu}
\email{yucong@mail.sysu.edu.cn}

\author[0000-0002-0447-7207]{Wei Zhong}
\affiliation{School of Physics and Astronomy, Sun Yat-Sen University, Zhuhai, 519082, People's Republic of China}
\affiliation{CSST Science Center for the Guangdong-Hong Kong-Macau Greater Bay Area, Zhuhai, 519082, People's Republic of China}
\affiliation{State Key Laboratory of Lunar and Planetary Sciences, Macau University of Science and Technology, Macau, People's Republic of China}

\author[0009-0004-1986-2185]{Zhen-Tai Zhang}
\affiliation{School of Physics and Astronomy, Sun Yat-Sen University, Zhuhai, 519082, People's Republic of China}
\affiliation{CSST Science Center for the Guangdong-Hong Kong-Macau Greater Bay Area, Zhuhai, 519082, People's Republic of China}
\affiliation{State Key Laboratory of Lunar and Planetary Sciences, Macau University of Science and Technology, Macau, People's Republic of China}

\author[0009-0005-9413-9840]{Hui-Sheng Zhong}
\affiliation{School of Physics and Astronomy, Sun Yat-Sen University, Zhuhai, 519082, People's Republic of China}
\affiliation{CSST Science Center for the Guangdong-Hong Kong-Macau Greater Bay Area, Zhuhai, 519082, People's Republic of China}

\author[0000-0002-0378-2023]{Bo Ma}
\affiliation{School of Physics and Astronomy, Sun Yat-Sen University, Zhuhai, 519082, People's Republic of China}
\affiliation{CSST Science Center for the Guangdong-Hong Kong-Macau Greater Bay Area, Zhuhai, 519082, People's Republic of China}

\author[0000-0003-2278-6932]{Xianyu Tan}
\affiliation{Tsung-Dao Lee Institute \& School of Physics and Astronomy, Shanghai Jiao Tong University, Shanghai 201210, China}

\author[0000-0003-0454-7890]{Cong Yu}
\affiliation{School of Physics and Astronomy, Sun Yat-Sen University, Zhuhai, 519082, People's Republic of China}
\affiliation{CSST Science Center for the Guangdong-Hong Kong-Macau Greater Bay Area, Zhuhai, 519082, People's Republic of China}
\affiliation{State Key Laboratory of Lunar and Planetary Sciences, Macau University of Science and Technology, Macau, People's Republic of China}

\begin{abstract}


Observations have revealed unique temperature profiles in  hot Jupiter atmospheres. We propose that the energy transport by vertical mixing could lead to such thermal features. In our new scenario, strong absorbers, TiO and VO are not necessary. 
Vertical mixing could be naturally excited by atmospheric circulation or internal gravity wave breaking. 
We perform radiative transfer calculations by taking into account the vertical mixing driven energy transport. The radiative equilibrium (RE) is replaced by radiative-mixing equilibrium (RME). We investigate how the mixing strength, $K_{\rm zz}$, affects the atmospheric temperature-pressure profile.
Strong mixing can heat the lower atmosphere and cool the upper atmosphere. This effect has important effects on the atmosphere thermal features
that would form without mixing.  In certain circumstances, it can induce temperature inversions in scenarios where  the temperature 
monotonically increases with increasing pressure under conditions of lower thermal band opacity.
Temperature inversions show up as $K_{\rm zz}$ increases with  pressure due to shear interaction with the convection layer.
The atmospheric thermal structure of HD~209458b can be well fitted with $K_{\rm zz} \propto (P/1\ {\rm bar})^{-1/2}\ {\rm cm}^{2} \ {\rm s}^{-1}$. 
Our findings suggest vertical mixing  promotes temperature inversions and lowers $K_{\rm zz}$ estimates compared to prior studies.
Incorporating chemical species into vertical mixing will significantly affect the thermal profile due to their temperature sensitivity.

\end{abstract}


\section{Introduction} \label{sec:intro}

Hot Jupiters orbit close to the host star and receive tremendous radiation from their parent star \citep{2007prpl...Charbonneau}. 
Due to the vicinity to their host star, they usually become tidally locked \citep{1996ApJ..Guillot}. 
This results in a significant temperature disparity between the day and night sides. 
Hot Jupiters constitute a distinct category among more than 5,700 known exoplanets, characterized by their expanded radii  (e.g., \citealt{2002A&A...Guillot,2004ApJ...Gu,2010ApJ...Batygin,2010ApJ...Youdin,2013ApJ...Spiegel,2017ApJ...Komacek,2022MNRAS...Hou}) and diverse infrared emission properties, some of which lead to stratospheric thermal inversions (\citealt{2003ApJ...Hubeny}; \citealt{2005ApJ...Fortney}; 
\citealt{2010A&A...Guillot}, hereafter G\citeyear{2010A&A...Guillot}).
Notably, the Spitzer Space Telescope detected this feature in the spectrum of the hot Jupiter HD~209458b \citep{2008ApJ...Knutson}.
In hot Jupiters, primary chemicals like Titanium oxide (TiO) and Vanadium oxide (VO) \citep{2003ApJ...Hubeny,2005ApJ...Fortney} absorb ultraviolet (UV)/optical light in the upper atmosphere, causing localized heating and temperature inversions. 
However, in cooler hot Jupiters like HD~209458b, these chemicals would condense deeper and be hard to detect, making temperature inversions unlikely \citep{2016ApJ...Line,2024MNRAS.Roth}.  Specifically, HD 209458b is currently not considered to have a temperature inversion \citep{2014ApJ...Diamond-Lowe,2016AJ....152..Line}. While many other planets, particularly ultra-hot Jupiters, exhibit clear temperature inversions, HD 209458b does not fall into this category.
\cite{2009ApJ..Spiegel} proposed that strong vertical mixing is necessary to maintain enough stratospheric TiO for  a thermal inversion.


Vertical mixing plays a crucial role in tracking the movement of chemicals and clouds \citep{2016AA...Fromang,2018MNRAS...Ryu,2018ApJ...Zhang....1Z,2018ApJ...Zhang,2019MNRAS...Menou,2019A&A...Ormel},
and contributes significantly to chemical disequilibrium \citep{2007ApJ...669.Hubeny,2022MNRAS.511.Tan,2023MNRAS.523.Lee} in the atmosphere.
Effective vertical mixing prevents the appearance of cold traps \citep{2009ApJ..Spiegel,2013A&A...Parmentier,2023Natur.Pelletier,2024arXivZhang}, and promotes both cloud formation and transport of materials from deeper atmospheric layers to higher altitudes \citep{2024arXivSing}.
Recent studies have determined the vertical mixing strength in the atmosphere of WASP-107b through spectroscopic observations and  atmospheric retrievals. 
\cite{2024arXivSing} used models incorporating vertical mixing and photochemistry, obtaining $\log_{10} \left(K_{\rm zz}/{\rm cm^2 \ s^{-1}}\right) = 11.6 \pm 0.1 $, demonstrating that vertical mixing triggers methane depletion.  
\cite{2024arXivWelbanks} included tidal effects, finding $ \log_{10} \left(K_{\rm zz}/{\rm cm^2 \ s^{-1}}\right)  = 8.4-9.0$ . 
These studies highlight the varying strengths and effects of vertical mixing between different models, making an accurate estimation of the vertical mixing strength crucial.

This work focuses on the energy transport of vertical mixing in a radiative atmosphere, contrasting with previous research that emphasized mass transport by turbulence and large-scale convection \citep{Zhang_2020}, and the interaction between horizontal mixing and radiative transfer (G\citeyear{2010A&A...Guillot}). In low-density upper atmospheres, such as Earth's mesosphere, breaking gravity waves significantly mix chemical tracers \citep{1987JGR....92.6691S, Zhang_2020}. \cite{1981JGR....86.9707L} first parameterized the eddy diffusion coefficient of these waves. In atmospheres with stable stratification, vertical mixing is driven by large-scale overturning and wave mixing \citep{1984maph...Holton, Zhang_2020}. However, the role of vertical mixing energy in the radiative layer, especially in conjunction with radiative transfer, remains unexplored. Investigating this energy transport is essential.

We study the influences of vertical mixing, driven by atmospheric circulation \citep{Zhang_2020} 
or breaking gravity waves \citep{1987JGR....92.6691S, Zhang_2020}
, on the temperature-pressure profiles of exoplanet atmospheres.
Actually, the energy transport induced by the vertical mixing provides an alternative explanation for the thermal inversion. This differs from the mechanisms proposed by \cite{2003ApJ...Hubeny,2008ApJ...Fortney,2018ApJ...Lothringer,2019MNRAS...Gandhi}, which attribute strong inversions to the presence of visible absorbers (TiO, VO, Fe, AlO, CaO, NaH, and MgH) with reasonable abundance or low infrared opacity due to the low abundance of ${\rm H_2O}$.
This study utilizes the mechanical greenhouse effect schematic as presented in \cite{2010ApJ...Youdin}. Unlike convection, which generates an outward energy flux, forced turbulence in stable regions induces a downward energy flux. This flux 
introduces changes to the planet's radiative equilibrium (RE), resulting in a radiative-mixing equilibrium (RME). 
This downward heat flux can compensate for or even exceed the convective losses.
The vertical mixing intensity $K_{\rm zz}$ will be treated as a free parameter. 
An increase in $K_{\rm zz}$ leads to stronger vertical mixing. 
Furthermore, the temperature variations are controlled by changes in the gradient of the mixing flux. 
As the depth increases, the gradient of the mixing flux first intensifies and then diminishes, resulting in the development of a temperature inversion.  
Since vertical mixing could heat the lower atmosphere, we can estimate the mixing intensity by matching the temperature-pressure profiles with those retrieved by transmission observations.

This work is structured as follows. The RME driven by vertical mixing is shown in \S \Ref{sec:atmosphere}. Additionally, we examine the impact of vertical mixing on the structure of the semi-grey atmosphere and verify its consistency with  
the atmospheric simulation results of \cite{2008ApJ..Showman} in \S \Ref{sec_reslut}. Finally, conclusions and discussions are presented in \S \ref{sec_conclusion}.

\section{The Irradiated Atmosphere} \label{sec:atmosphere}

A planetary atmosphere that is affected by external radiation, such as that from a star, is referred to as an irradiated atmosphere.
Atmospheric circulation or breaking gravity waves can create vertical mixing and force downward flux, similar to how a greenhouse works, possibly changing the temperature structures in planetary \citep{2010ApJ...Youdin}.
We here set up the theoretical framework used to examine how vertical mixing can affect semi-grey atmospheres. 
We summarize the equations related to radiative equilibrium in \S \ref{sec:Radiative_transfer}, and the equations related to RME in \S \ref{sec:RTE}. 
By solving these equations, the effects of the vertical mixing on the exoplanetary atmosphere could be investigated.

\subsection{Radiative Transfer}\label{sec:Radiative_transfer}

In a static plane-parallel atmosphere under local thermodynamic equilibrium, the moments of the radiative transfer equation (RTE; \citealt{1960book.Chandrasekhar,1978book..Mihalas}; G\citeyear{2010A&A...Guillot}) can be simplified as follows:
\begin{equation}
\frac{\mathrm{d} H_{\nu}}{\mathrm{d} \tau_{\nu}} = J_{\nu} - B_{\nu} \ ,
\label{eq_H}
\end{equation}
\begin{equation}
\frac{\mathrm{d} K_{\nu}}{\mathrm{d} \tau_{\nu}} = H_{\nu} \ ,
\label{eq_K}
\end{equation}
without considering the effect of scattering\footnote{The interplay between the scattering (both isotropic and anisotropic) and vertical mixing would be reported elsewhere.}.
Here, $B_{\nu}$ is the source function and $J_{\nu}$, $H_{\nu}$, and $K_{\nu}$ are the zeroth, first, and second radiation moments, all dependent on frequency $\nu$. 
Unlike \cite{2017MNRAS...GENESIS}, which uses opacity per unit length $\kappa_{\nu}'$, we use opacity per unit mass, defined as $\kappa_{\nu} = \kappa_{\nu}'/\rho$ as in G\citeyear{2010A&A...Guillot}.
The optical depth relationship is given by $d\tau_{\nu} = - \kappa_{\nu} dz$ with altitude $z$.
The second-order radiative transfer equation is simplified as \citep{2017MNRAS...GENESIS,2017MNRAS...Hubeny}
\begin{equation}
 \frac{\partial^2 K_{\nu}}{\partial \tau_\nu^2} = \frac{\partial^2\left(f_\nu J_{\nu}\right)}{\partial \tau_\nu^2} 
 = J_{\nu}-B_{\nu} \ ,
     \label{eq:RTE}
\end{equation}
with the Eddington factor  $f_\nu = K_\nu/J_\nu$.

The atmospheric boundaries \citep{2017MNRAS...GENESIS} at the top and bottom are defined as follows:
\begin{equation}
     \left.\frac{\partial\left(f_{\nu} J_{\nu}\right)}{\partial \tau_{\nu}}\right|_{\tau_\nu=0}= g_\nu J_{\nu 0}-H_{\mathrm{ext}}\ ,
     \label{eq:upper_boundary}
\end{equation}
\begin{equation}
     \left.\frac{\partial\left(f_{\nu} J_{\nu}\right)}{\partial \tau_{\nu}}\right|_{\tau_{\nu}=\tau_{\max }}=\left[\frac{1}{2}\left(B_{\nu}-J_{\nu}\right)+\frac{1}{3} \frac{\partial B_{\nu}}{\partial \tau_\nu}\right]_{\tau_{\nu} = \tau_{\max}} \ .
\label{eq:bottom_boundary}
\end{equation}
Eq.(\ref{eq:bottom_boundary}), known as the diffusion approximation, highlights the significance of optical thickness with high opacity and gas density. 
The coefficient $g_{\nu} = H_{\nu 0}/J_{\nu 0}$ represents the ratio of incoming to outgoing radiation at $\tau_{\nu}=0$. 
Additionally, external stellar irradiation, denoted as $H_{\rm ext}$, can be calculated from the incoming flux $F_{\rm ext}$ using $H_{\rm ext} = F_{\rm ext}/4\pi$.

In the atmosphere RE \citep{2017MNRAS...Hubeny} could be expressed as:
\begin{equation}
     \alpha \int_0^{\infty} \kappa_\nu(J_\nu -B_\nu) d \nu + \beta \int_0^{\infty} \frac{d(f_{\nu} J_{\nu})}{d \tau_{\nu}} d \nu = \frac{\sigma_{\rm R}}{4 \pi} T_{\mathrm{int}}^4 \ ,
     \label{eq:conservation_case01}
\end{equation}
with the Stefan-Boltzmann constant $\sigma_{\rm R}$ and the intrinsic temperature $T_{\rm int}$. 
The RTE tends to diverge and must be regulated using Eq.(\ref{eq:conservation_case01}). 
The critical threshold is established at a pressure of $P_{\rm crit}= 1\, \text{bar}$. 
When the pressure is below $P_{\rm crit}$, the coefficients $\alpha$ and $\beta$ are assigned values of 0 and 1, respectively. 
In contrast, when pressure exceeds $P_{\rm crit}$, the coefficients $\alpha$ and $\beta$ are set to 1 and 0, respectively. 
This method effectively reduces numerical instabilities caused by high opacity.

\subsection{Radiative-Mixing Equilibrium} \label{sec:RTE}

Under typical circumstances and in the absence of external influences, heat is transferred from warmer to cooler areas in accordance with the second law of thermodynamics. 
Nevertheless, close to the IR photosphere, vertical mixing can occur due to atmospheric circulation or the breaking of gravity waves. 
This phenomenon enables heat to move from cooler to warmer regions, thus disrupting the conventional heat transfer routes and modifying the energy balance of the system.

Since the sound wave travels faster than the heat diffuses in the fluid parcel, 
it conserves entropy $S$ and maintains the equilibrium of pressure $P$ with its surroundings as it moves radially over a distance $\ell$. 
The temperature disparity between the blob and its environment is 
\begin{equation}
\delta T = \left(\left.\frac{dT}{dz}\right|_{\mathrm{ad}} - \frac{dT}{dz}\right) \ell = - \ell \frac{ T}{c_{\rm p} } \frac{dS}{dz} \ , 
\end{equation}
with isobaric heat capacity $c_{\rm p}$. 
In a stably stratified region, the entropy gradient is given by $dS/dz = g/T\left(1-\nabla/\nabla_{\rm ad}\right)>0$. The descending particles have the potential to warm the atmosphere.
The extra heat $\delta q = \rho c_{\rm p} \delta T$. Thus, the heat flux by vertical mixing flux is given by 
\begin{equation}
F_{\text{mix}} = w \delta q = -K_{\rm zz} \rho g \left(1 - \frac{\nabla}{\nabla_{\mathrm{ad}}}\right) \ ,
\label{eq_f_mix}
\end{equation}
with the eddy speed $w$,  the mixing diffusion $ K_{\rm zz} = w \ell $, and gravitational acceleration $g$.
While $\nabla_{\mathrm{ad}} = 2/7$ indicates the adiabatic temperature gradient, $ \nabla = \delta \ln T/ \delta \ln P$ is for radiative layer 
\citep{2014tsa..book...Hubeny,2017MNRAS...Hubeny}.

The temperature distribution is governed by the energy equilibrium. 
In areas where radiation is the primary mode of energy transfer, a perfect equilibrium between incoming and outgoing radiation can be maintained \citep{2017MNRAS...GENESIS}. 
However, when the vertical mixing within the planet is considered, the usual RE for the upper atmosphere is replaced by RME, which can be written as
\begin{equation}
     \int_0^{\infty} \kappa_\nu \left( J_\nu-B_\nu \right) \mathrm{d} \nu +\frac{ g}{4 \pi} \frac{\mathrm{d} F_{\text {mix}}}{\mathrm{d} P}=0\ .
     \label{eq_rme1}
\end{equation}
However, Eq. (\ref{eq_rme1})  does not prevent divergence in the deeper layers of the simulation. 
Given that $dP = g dm$ and $d\tau = \kappa_{\rm R} dm$, the RME Eq. (\ref{eq_rme1}) can be reformulated by integrating Eq. (\ref{eq_H}) into a new conservation equation for the total flux in the deeper atmosphere, i.e.,
\begin{equation}
     \int_0^{\infty} \frac{\mathrm{d}\left(f_\nu J_\nu\right)}{\mathrm{d} \tau_{\nu}} \mathrm{~d} \nu+\frac{F_{\text {mix}}}{4 \pi}=\frac{\sigma_{\rm R}}{4 \pi} T_{\text {int }}^4 \ .
     \label{eq_rme2}
\end{equation}
Density is derived by the ideal gas equation $P = {\rho k_{\mathrm{b}} T}/{\overline{m}}$, incorporating the mean molecular mass $\overline{m}$ and Boltzmann constant $k_{\mathrm{b}}$. 
The radiative transfer equations become nonlinear when the RME constraint is to be satisfied. 
We use Rybicki's method \citep{2014tsa..book...Hubeny,2017MNRAS...Hubeny,2019PhDT...Gandhi} to solve 
the nonlinear Eqs. (\ref{eq:RTE}-\ref{eq:bottom_boundary}) and (\ref{eq_rme1}-\ref{eq_rme2}). 
The Rybicki scheme starts with the same set of linearized structural equations, reorganizing the state vector and the Jacobi matrix. Solutions are obtained by solving the tridiagonal matrices using Newton-Raphson iteration.
Refer to \S 3.3 in \citet{2017MNRAS...Hubeny} or \S 17.3 in \citet{2014tsa..book...Hubeny}.
By combining it with the setup in Eqs. (\ref{eq:11}-\ref{eq:13}) from \S \ref{sec3.1}, we can gradually approach the correct $T$-$\tau$ profiles driven by vertical mixing.

\section{Results}\label{sec_reslut}

When vertical mixing is taken into account, the atmospheric structure would be altered accordingly. In \S \ref{sec3.1}, we elaborate on how the vertical mixing flux can change atmospheres. 
The development of temperature inversions in monotonically increasing temperature profiles is shown in \S \ref{sec3.2}. 
Furthermore, our results in \S \ref{sec3.3} demonstrate that mixing aids in better alignment between the semi-grey atmospheric model of G\citeyear{2010A&A...Guillot} and the atmospheric structure of HD~209458b from \cite{2008ApJ..Showman}.

\subsection{The Effects of Vertical Mixing}
\label{sec3.1}

Here we describe how we incorporate the mixing induced energy transport into the calculation of the atmosphere temperature profile.
In a semi-grey atmosphere, where opacity arises from both visible and thermal infrared bands, the energy equation corresponding to Eq.(\ref{eq_rme1}) at the upper atmosphere simplifies to 
\begin{equation}
    \kappa_{\rm th} \left(J_{\rm th} - B\right) + \kappa_{\rm v}J_{\rm v} +\frac{
    g}{4 \pi} \frac{\mathrm{d} F_{\text {mix}}}{\mathrm{d} P}=0\ .
\end{equation}
Variables with the subscript ``${\rm v}$'' denote the visible (or optical) band, while those with the subscript ``${\rm th}$'' relate to the thermal (or infrared) band.
The temperature is determined by the value of $B$, which must satisfy the following condition:
\begin{equation}
 B = J_{\rm th} + \frac{1}{\kappa_{\rm th}} \left(\frac{ 
 g}{4\pi} \frac{d F_{\rm mix}}{d P} +\kappa_{\mathrm{v}} J_{\mathrm{v}} \right) \ .
 \label{eq_B1}
\end{equation}
The variation of $F_{\rm mix}$ with respect to pressure is:
\begin{equation}
    \frac{\mathrm{d} F_{\text {mix}}}{\mathrm{d} P} 
    =  \frac{F_{\rm mix}}{P}\left(1-\nabla\right) +  \frac{K_{\rm zz}\rho g}{\nabla_{\rm ad}} \frac{d\nabla}{d P} + F_{\rm mix}\frac{d \ln K_{\rm zz}}{dP}\ ,
\label{eq_det}
\end{equation}
with $\rho = \overline{m}  P/k_{\rm B}T$ and $\nabla = d \ln T/d\ln P$. 
In this section, we focus on a spatially constant $K_{\rm zz}$. Under this condition, the third term on the right side of Eq.(\ref{eq_det}) can be ignored.
Vertical mixing does not alter the first three moments of radiation intensity in the visible spectrum. This is expressed as 
\begin{equation} 
\left[H_{\rm v}, J_{\rm v} \right] = \left[H_{\rm v0} , \ J_{\rm v0} \right]{\rm e}^{-\gamma \tau / \mu} \ , 
\end{equation} 
with $\gamma=\kappa_{\rm v}/\kappa_{\rm th}$ \citep{2010ApJ...Youdin}. These moments only change with the optical depth.

The temperature at the top atmosphere $T_0$ can also be modified. Given that the atmospheric temperature profile is determined by Eq.(\ref{eq_B1}), $T_0$ at the top boundary conforms to this relationship
\begin{equation}
T_{\rm 0} \propto B_{0}^{1/4}  \approx  \left[\frac{H_{\rm th0}}{g_{\rm th}} + \frac{1}{\kappa_{\rm th}} \frac{
g}{4\pi} \frac{ F_{\rm mix0}}{ P_{\rm 0}} +\frac{\kappa_{\mathrm{v}}}{\kappa_{\rm th}} J_{\mathrm{v0}} \right]^{1/4} \ ,
\label{eq_sum}
\end{equation}
assuming $d \nabla / d P \approx 0$ and $\nabla$ is significantly smaller at this boundary. 
The values of all parameters are known, except for the term containing $F_{\rm mix}$ and $J_{\rm th}$. 
Furthermore, because $P$ is proportional to $\tau$, it can be assumed constant at the top. 
The very low density at the top $\rho_{\rm 0}$ has minimal influence on other parameters and can be considered constant.
Since the RME of Eq.(10) in the semi-grey atmosphere should be
\begin{equation}
    H_{\rm v} + H_{\rm th} + \frac{F_{\rm mix}}{4\pi} =\frac{\sigma_{\rm R}}{4 \pi} T_{\text {int }}^4 \ ,
\end{equation}
we can determine the value of $H_{\rm th0}$ at top boundary as shown below
\begin{equation}
    H_{\rm th0} = \frac{\sigma_{\rm R}}{4 \pi} T_{\text {int }}^4  - H_{\rm v0} - \frac{F_{\rm mix0}}{4\pi} = c_{\rm 1} - \frac{F_{\rm mix0}}{4\pi} \ .
\label{eq_outer H}
\end{equation}
Here $c_{\rm 1}$ is a constant, defined as $c_{\rm 1} = {\sigma_{\rm R}}/{4 \pi} T_{\text {int }}^4 - H_{\rm v0}$.
Combining Eq.(\ref{eq_outer H}) and Eq.(\ref{eq_sum}), the final expression for $T_{\rm 0}$ becomes:
\begin{equation}
\begin{aligned}
    T_{\rm 0} \propto &  \left[\left(-\frac{1}{4 \pi g_{\rm th}} + \frac{1}{\kappa_{\rm th}} \frac{  g}{4\pi  P_{\rm 0}} \right)F_{\rm mix0} +\frac{\kappa_{\mathrm{v}}}{\kappa_{\rm th}} J_{\mathrm{v0}} + \frac{c_{\rm 1}}{g_{\rm th}}\right]^{1/4}   \\
    \propto & \left[c_{\rm 3} \  F_{\rm mix0} + c_{\rm 2} \right]^{1/4} \ .
\end{aligned}
\label{eq_oouter_t0}
\end{equation}
The parameters $P_{\rm 0}$, $c_{\rm 2}$, and $c_{\rm 3}$ are also constant at the top boundary. 
The last two are defined as $c_{\rm 2} = {\kappa_{\mathrm{v}}}J_{\mathrm{v0}}/{\kappa_{\rm th}}  + {c_{\rm 1}}/{g_{\rm th}}$ and   $c_{\rm 3} = \left(- 1/g_{\rm th} + g/ \kappa_{\rm th} P_{\rm 0}\right)/4 \pi$, respectively.
As a result, the change in $T_{\rm 0}$ depends on $F_{\rm mix0}$. 

\begin{figure*}
    \centering
    \includegraphics[width=1.0\linewidth]{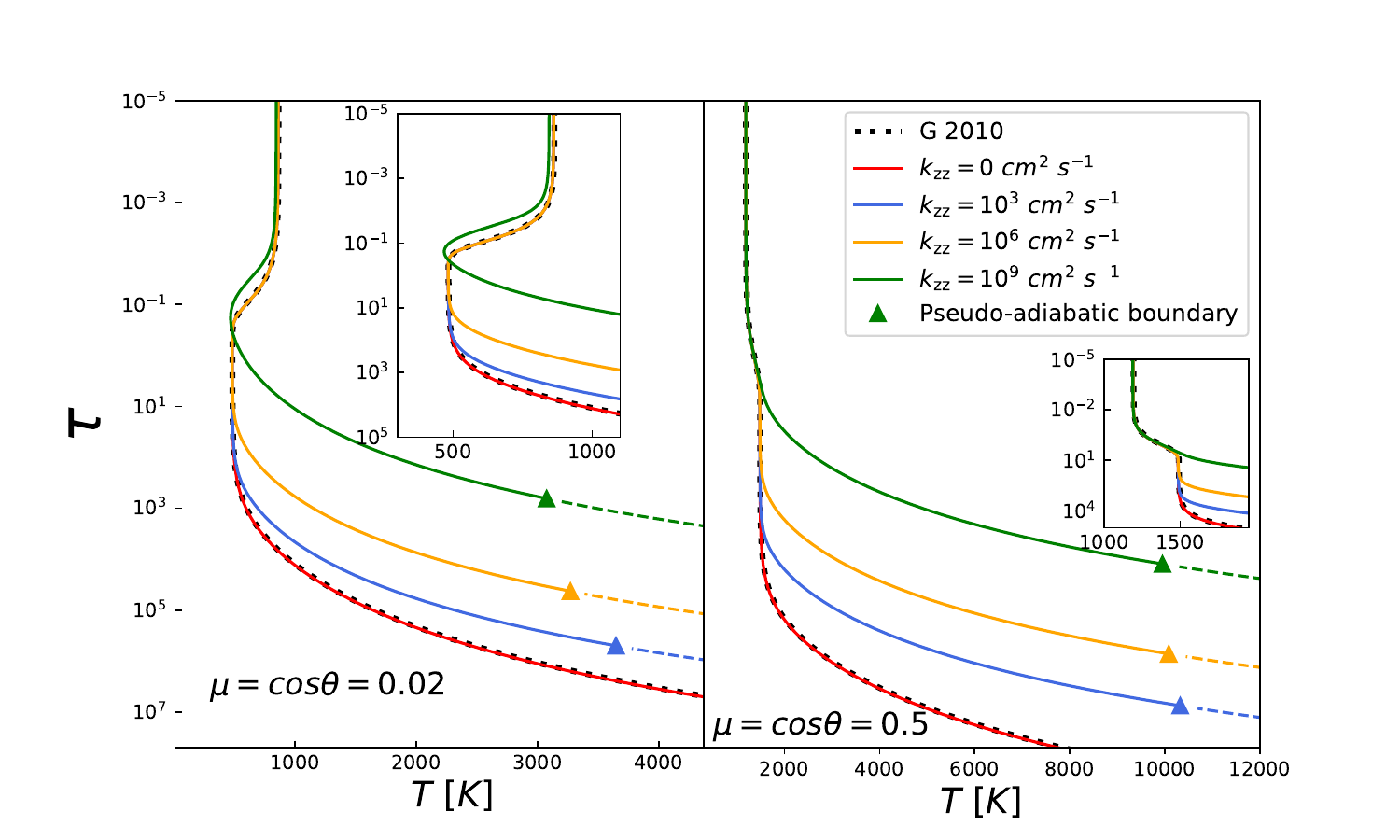}
    \caption{The left panel shows the effect of varying mixing strengths $K_{\rm zz}$ on the semi-grey atmosphere for $\mu = 0.02$ as $\nabla_{\rm ad}=2/7$, whereas the right panel depicts the scenario for $\mu = 0.5$. The small panels within the left and right panels are zoomed-in views of local areas. The colors red, blue, orange, and green correspond to increasing $K_{\rm zz}$. The black dots represent the atmospheric structure obtained from Equation (27) in G2010. 
    The onset of the pseudo-adiabatic zone's formation is denoted by   the filled triangle.
   The dashed lines denote the pseudo-adiabatic regions.
    Additionally, $g = 1000 \ \rm cm \ s^{-2}$, $T_{\rm int} = 100\ K$, and $T_{\rm irr} = 1469\ K$. }
    \label{fig1_modifies}
\end{figure*}

\begin{figure*}
    \centering
    \includegraphics[width=1.0\linewidth]{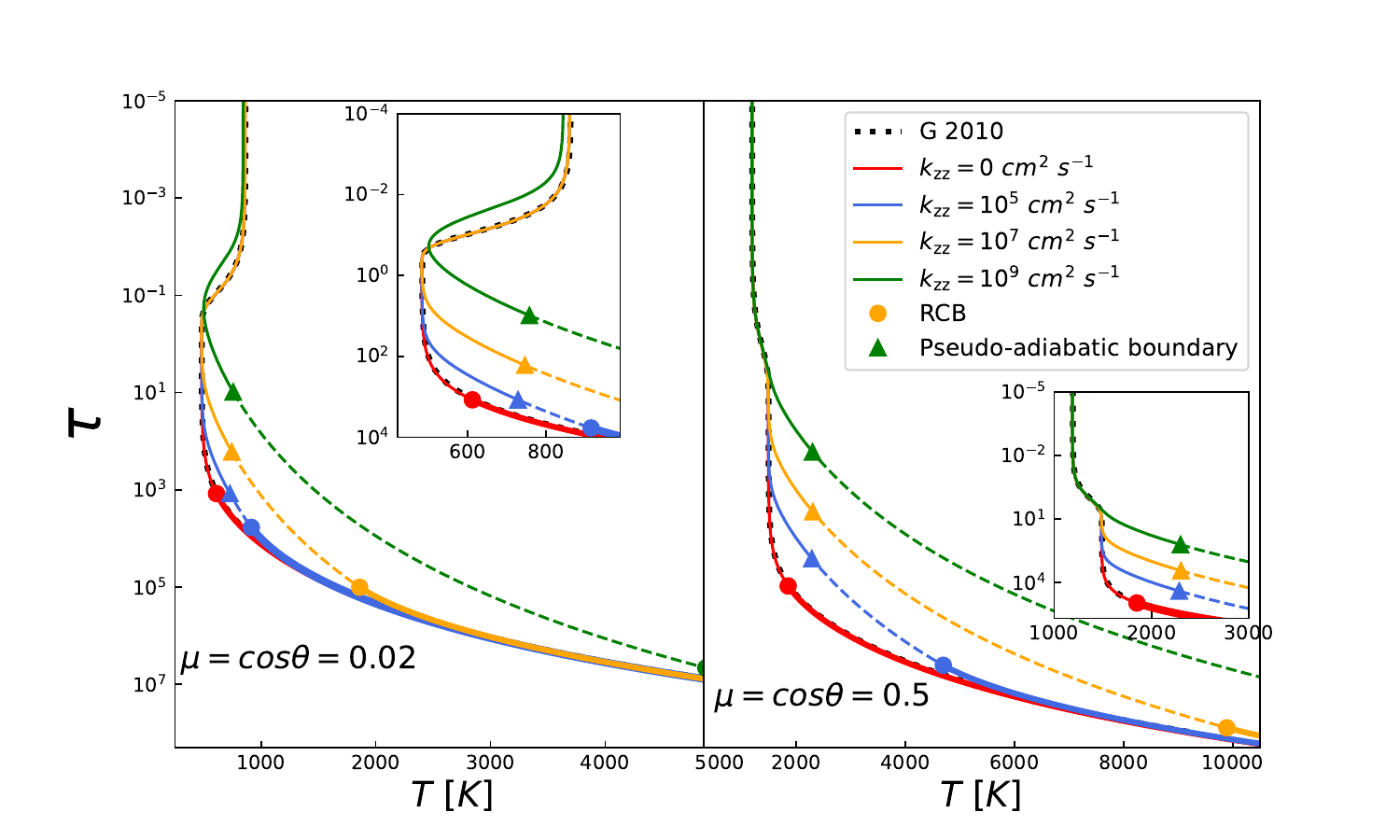}
    \caption{The left panel shows the effect of varying mixing strengths $K_{\rm zz}$ on the semi-grey atmosphere for $\mu = 0.02$ as $\nabla_{\rm ad}=1/7$, whereas the right panel depicts the scenario for $\mu = 0.5$. 
    The mixing strength $K_{\rm zz}$ ranges from 0 to $10^9\ cm^2/s$, represented by different colors.
    The triangular markers denote the boundaries of the pseudo-adiabatic region, and the circular markers indicate the radiation-convection boundaries (RCB).
    The dashed lines denote the pseudo-adiabatic regions, while the thick solid lines indicate the convectively unstable region.
    All other parameters are consistent with Figure \ref{fig1_modifies}. }
    \label{fig1_modified_02}
\end{figure*}

The structure of planetary atmospheres is determined by Eqs.(\ref{eq_B1}) and (\ref{eq_oouter_t0}). However, the mixing intensity term is non-linear, so we need to validate our results by comparing numerical simulations with the analytical solutions of these equations.
We employ a semi-grey atmospheric model that is capable of both transporting inherent heat and receiving external irradiation.
We follow the numerical method proposed by \cite{2017MNRAS...GENESIS} to recalculate the atmospheric thermal structures, treating the opacity in the visible and thermal bands in the same way as in G2010. 
The opacity values for the visible and thermal bands are $\kappa_{\mathrm{v}} = 4 \times 10^{-3} \mathrm{~g~cm}^{-2}$ and $\kappa_{\mathrm{th}} = 10^{-2} \mathrm{~g~cm}^{-2}$, respectively. 
 The value of $\gamma = \kappa_{\rm v}/\kappa_{\rm th}$ shown in Figures~\ref{fig1_modifies} and \ref{fig1_modified_02} is identical.
Additionally, the top boundary conditions satisfy:
\begin{equation}
    \left(H_{\rm v0}, J_{\rm v0}, H_{\rm ext}\right) = \left(-\mu J_{\rm v0}, \frac{\sigma T_{\rm irr}^4}{4 \pi}, \mu \frac{\sigma T_{\rm irr}^4}{4 \pi} \right).
    \label{eq:11}
\end{equation}
The Eddington factors for the visible and thermal bands should satisfy:
\begin{equation}
    f_{\rm v} = \frac{K_{\rm v}}{J_{\rm v}} = \mu^2 \ ,  \quad g_{\rm v} = \frac{H_{\rm v}}{J_{\rm v}} = -\mu \ ,
    \label{eq:12}
\end{equation}
\begin{equation}
    f_{\rm th} = \frac{K_{\rm th}}{J_{\rm th}} = \frac{1}{3} \ , \quad g_{\rm th} = \frac{H_{\rm th0}}{J_{\rm th0}} = \frac{1}{2} \ ,
    \label{eq:13}
\end{equation}
where the incident angle  cosine $\mu = (K_{\rm{v}} / J_{\rm{v}})^{1/2}$. In particular, the Eddington factor $g_{\rm v}$ is zero at the top boundary.
Moreover, the Planck function integrated over all frequencies in the thermal range is given by $B = \int_{0}^{\infty} B_{v} dv = \sigma T^4/\pi$ because $B_{\rm v} = 0$ in the visible spectrum. 

We will apply the method outlined in \S 2.2, along with the above parameters, to solve the non-linear equations of vertical mixing cases. 
Vertical mixing can affect the formation of a temperature inversion. \cite{2009ApJ..Spiegel} estimated the mixing strength $K_{\rm zz} \approx 10^7 - 10^{11} \, {\rm cm}^2 \, {\rm s}^{-1}$ to ensure enough TiO in the stratosphere for a thermal inversion.  Recent constraints on $K_{\rm zz}$ inferred from the JWST spectrum of WASP-107b have positioned $K_{\rm zz}$ at a markedly elevated level, as exemplified by $10^{8.4}-10^{9}\ cm^2/s$ \citep{2024arXivWelbanks}. Nevertheless, the presence of a deep radiative zone \citep{2023NatAs..Cavali} may mitigate the constraint of $K_{\rm zz}$. Consequently, we opt for a reduced and more conservative estimate of $K_{\rm zz} \approx 10^{9} cm^2/s$ for the sample. Here, we consider $K_{\rm zz} \approx 0-10^9 \, {\rm cm}^2 \, {\rm s}^{-1}$. 
Strong mixing could potentially exist if we do not address an upper limit of $K_{\rm zz}$ in order for the planet to maintain its observed radius and avoid overinflation \citep{2010ApJ...Youdin}. 
The results are presented as the different lines in Figure~\ref{fig1_modifies} and Figure~\ref{fig1_modified_02}.
The outcomes of the RTE simulation, governed by the RE conditions, align with the analytical profile from G\citeyear{2010A&A...Guillot}. 
The alignment is shown by the red line for the non-mixing case coinciding with the black dotted line, demonstrating the simulation's precision.

Radiative transfer focuses on atmospheric opacity, which is influenced by its chemical composition across different spectral bands. 
As in $\kappa_{\rm v}<\kappa_{\rm th}$, visible radiation passes through more easily than heat radiation, trapping heat and causing a sustained increase in temperature. 
In contrast, a higher $\kappa_{\rm v}$ blocks more visible radiation, leading to thermal inversion by trapping heat higher up and preventing cooling \citep{2014A&A...Parmentier}. 
However, changes in cosine of the incident irradiation angle $\mu = \cos \theta$ can alternate these signatures, as shown by the black dotted line in Figure~\ref{fig1_modifies} and Figure~\ref{fig1_modified_02}, derived by Eq.(27) from G\citeyear{2010A&A...Guillot}. 
A lower value of $\mu$ can be associated with a thermal inversion signature, whereas a higher value of $\mu$ leads to a monotonic increase in temperature.

Figure~\ref{fig1_modifies} provides a comparison of the temperature structures obtained for diﬀerent values of the mixing strength $K_{\rm zz}$ and in the different cosine of incident angle $\mu$, as the adiabatic temperature gradient $\nabla_{\rm ad}=2/7$. 
When $\mu<0.5$, a temperature inversion occurs without mixing.  
Increasing the angle of incidence causes irradiation to be absorbed at higher altitudes, raising temperatures at lower pressures and forming a temperature inversion.
However, mixing reduces the width of the inversion region.
In this study, vertical mixing introduces a downward energy flux. 
This flux modifies the temperature profiles as in Eq.(\ref{eq_B1}) and Eq.(\ref{eq_oouter_t0}). At top atmosphere, the analytical derivation in Equation (\ref{eq_oouter_t0}) indicates that this flux theoretically lowers the temperature, but this effect becomes clear with stronger mixing.

In the upper atmosphere, the mixing flux $F_{\rm mix} \sim - \rho g K_{\rm zz} \propto P $ occurs in the regions where $\nabla\ll 1$. 
The temperature changes at deeper altitudes follow Eq.(\ref{eq_B1}) where variations in the gradient of $\left|F_{\rm mix}\right|$ dominate. 
The solid green line in the left panel indicates that strong mixing results in a cooler temperature  at low pressures than the no-mixing scenario shown by the solid red line. 
Moreover, the upper boundary for the inversion zone will occur at a comparatively shallower depth.
The temperature continues to decrease, in comparison with the non-mixing scenario, until the gradient of $\left|F_{\rm mix}\right|$ starts to diminish. At this point, the temperature starts to rise, leading to a narrowing of the inversion area. This can also trigger the  pseudo-adiabatic regions (as shown by dashed lines) due to increased temperature gradients. 

The presence of the pseudo-adiabatic region aligns with the findings of \cite{2010ApJ...Youdin} and \cite{2017ApJ...Yu}. 
Such regions are notably prevalent in the atmospheres of hot Jupiter-like exoplanets subjected to intense stellar radiation. 
A pseudo-adiabatic region has an internal temperature gradient slightly lower than the theoretical adiabatic gradient.
Furthermore, an enhancement in mixing strength might prompt the pseudo-adiabatic boundary to establish at shallower depths.
Concurrently, it is anticipated that the presence of such regions may drive the radiation-convection boundary (RCB) to extend deeper.

Weaker mixing aligns the upper temperature structure with the non-mixing case but still causes a temperature rise at shallower depths, i.e., pseudo-adiabatic regions form. The inversion region widens with decreased mixing intensity, but remains narrower than without mixing.
For $\mu \ge 0.5$, in the absence of mixing, the temperature increases monotonically  with increasing pressure. Vertical mixing raises the temperature of the lower atmosphere, making  pseudo-adiabatic regions more likely. 
Although intense mixing cools the upper atmosphere, it is insufficient to overcome opacity and create a temperature inversion zone within this parameter range.

For $\nabla_{\rm ad} = 2/7$, the unstable convection region cannot be triggered in any temperature profiles. As mentioned earlier, the emergence of a pseudo-adiabatic zone has the potential to displace the location of the RCB to greater depths. In order to substantiate our hypothesis, it is imperative to examine the various $\nabla_{\rm ad}$ that influence the positioning of the RCB. 
The temperature gradient is governed by the chemical composition. Nevertheless, this study employs a semi-grey model for thermal and optical spectra that excludes specific chemical influences, thus rendering the adiabatic temperature gradient indeterminate. 
The actual adiabatic temperature gradient for exoplanets may differ. 
To address this, we follow \cite{2015A&A...Parmentier} using variable multipliers for the adiabatic temperature gradient. In this section, we do not take into account the actual convective flux as described by \cite{2017MNRAS...GENESIS}. Instead, we employ the Schwarzschild criterion to determine the unstable convection boundary.

Figure~\ref{fig1_modified_02} presents the updated temperature profiles corresponding to varying mixing intensities, alongside the adiabatic temperature gradient as indicated in $\nabla_{\rm ad}=1/7$. In this figure, the strengths of vertical mixing are denoted as references $0$, $10^5$, $10^7$, and $10^9\ cm^2/s$.
The findings reveal that even with the reduced adiabatic temperature gradient, vertical mixing can substantially increase the deep temperatures of the planet, with a more pronounced effect as the mixing intensity increases. 
The alterations in the temperature structure conform to Eq. (\ref{eq_B1}) and Eq. (\ref{eq_oouter_t0}), exhibiting a similar trend as observed in the scenario described in $\nabla_{\rm ad}=2/7$.
The resultant temperature profile is anticipated to exhibit an elevation.
However, the efficiency of the heating is reduced in comparison to that described in $\nabla_{\rm ad}=2/7$.
Moreover, the pseudo-adiabatic region forms earlier under these conditions. 
Furthermore, with stronger mixing, the region forms at a shallower depth.
We observed that the formation of the pseudo-adiabatic region has, to some extent, pushed the RCB further inward. 
Greater mixing intensity leads to more significant inward movement.
Furthermore, the RCB may not be identifiable in the context presented in $\mu=0.5$.

 
 The temperatures of the deep atmosphere increase at relatively low optical depths, as noted by $\tau \sim 1$ for the scenario addressed in $K_{\rm zz} = 10^9\ cm^2 \ s^{-1}$. However, the overarching trend of temperature continues to increase towards higher optical depths, as illustrated in Figure~\ref{fig1_modifies} and Figure~\ref{fig1_modified_02}.
 In addition, we also examined the relationship between pressure and temperature. At $\tau \sim 1$, the pressure was found to be around $10^{-1}$ bar, which aligns well with the temperature increase trend observed for Uranus in Figure 4 of \cite{Zhang_2020}  and for thick-atmosphere worlds in the Solar System shown in Figure 1 of \citet{Robinson2014Nat}. 
Additionally, rising temperatures around the $\tau \sim 1$ will likely lead to the formation of a pseudo-adiabatic region. This seems plausible to us.

\begin{figure} 
     \centering
     \includegraphics[width=1.0\linewidth]{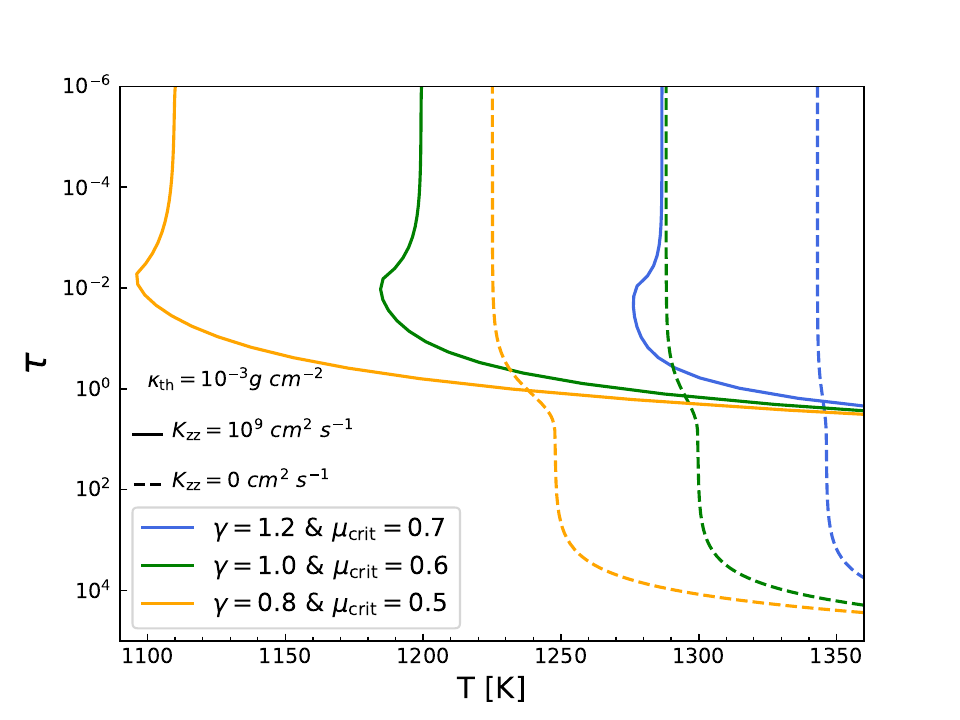}
     \caption{ 
     Formation of temperature inversion in monotonically increasing temperature profiles through vertical mixing under different critical cases. 
     Solid lines illustrate cases with vertical mixing, while dashed lines represent conditions without mixing. 
     The yellow, green and blue lines correspond to  $\gamma = $ 0.8, 1.0, and 1.2, respectively.
     }
     \label{fig2}
\end{figure}
\subsection{The Behavior of Monotonically Increasing Temperature Profiles with Vertical Mixing}
\label{sec3.2}

 In the absence of vertical mixing, the optical-temperature profile depends on factors such as opacity in two spectral bands, external irradiation, intrinsic temperature, and the cosine of the incident angle, as described by the semi-grey model. 
Typically, temperture inversion is anticipated when visible opacity exceeds thermal opacity. Furthermore, according to $\gamma=\kappa_{\rm v}/\kappa_{\rm th}=1.0$, the temperature profile is expected to be isothermal in such scenarios. However, the cosine of the incident angle $\mu$ will change this signature 
when other parameters remain constant. This effect is illustrated in Figures~\ref{fig1_modifies} and ~\ref{fig1_modified_02}, where $\kappa_{\rm v}<\kappa_{\rm th}$. When $\gamma \ge 1$, changes in the cosine of the incident angle affect the temperature structure in a similar way.  
This phenomenon stems from the influence of the angle of incidence on the absorption of radiative flux, i.e., $H_{\rm ext} = \mu \sigma T_{\rm irr}^4/4\pi$. 
Depicted in Figure~\ref{fig2}, there exists a critical parameter, such as $\mu_{\rm crit}$, which represents a critical value for the occurrence of a monotonic increase in the temperature profile. 
The results for our model, the $\mu_{\rm crit}$ increase with  $\gamma$.

Figure~\ref{fig2} demonstrates how monotonically increasing temperature profiles are influenced by vertical mixing. In cases of strong mixing, the resulting temperature inversions are relatively minor. Although inversions do form, their prominence decreases as $\gamma$ increases. Compared to the thermal opacity seen in the right panels of Figure~\ref{fig1_modifies} and Figure~\ref{fig1_modified_02}, the thermal opacity in this figure has been reduced by a factor of 10, which allows the inversion to occur. This suggests that in thicker envelopes, vertical mixing becomes less effective as opacity increases. While temperature inversions are an intriguing consequence of vertical mixing, they play a relatively small role. The most significant observational impact of vertical mixing is the increase in deep temperature.

\begin{figure} 
     \includegraphics[width=1.0\linewidth]{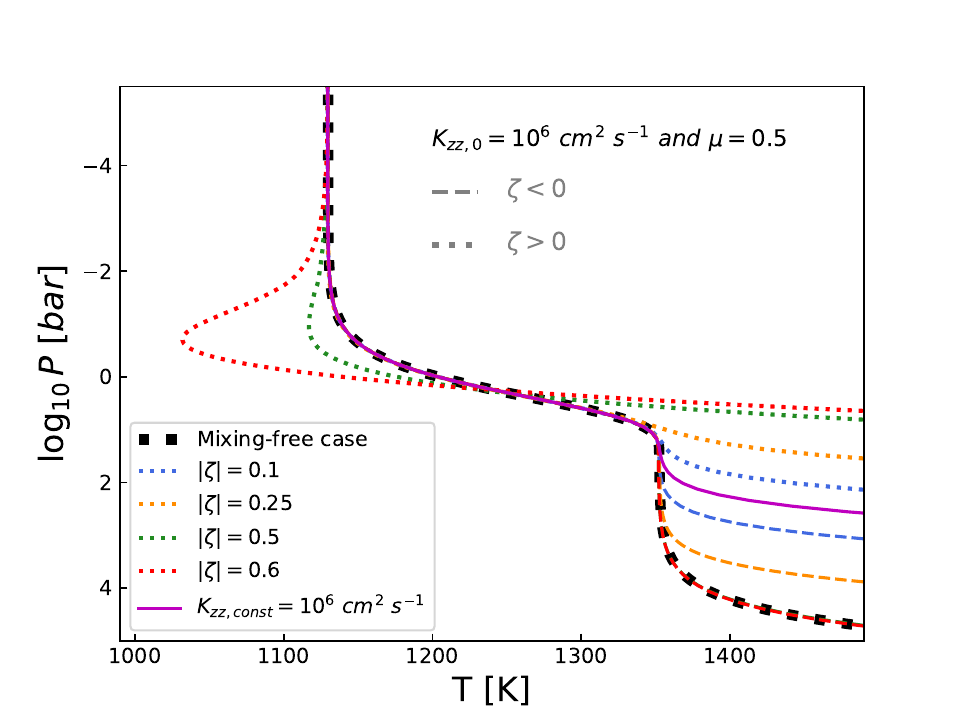}
     \caption{
    The temperature-pressure profiles in the atmosphere vary with the mixing coefficient $ K_{\rm zz}$ scaling as $ K_{\rm zz} \propto \left(P/ 1\ bar\right)^{\zeta} $.  The thick black dotted represents the scenario without mixing. Dotted and dashed lines indicate increased (positive $\zeta$) and decreased (negative $\zeta$) mixing strengths, respectively.  The colors red, green, orange, and royal blue correspond to $\left|\zeta\right|$ values of 0.6, 0.5, 0.25, and 0.1. Additionally, the purple line illustrates the scenario with a constant $ K_{\rm zz} $.
     }
     \label{fig3}
\end{figure}

In planetary atmospheres, the strength of vertical mixing varies with altitude due to dynamic conditions. 
We now further test the situation with a spatially variable mixing coefficient, $K_{\rm zz}$, defined as $K_{\rm zz} \simeq 10^6 \left(P/1 \text{bar}\right)^{\zeta} \ cm^2 \ s^{-1}$, which significantly impacts temperature profiles (see Figure \ref{fig3}).  When $\zeta$ is positive, $K_{\rm zz}$ increases with increasing pressure, and when $\zeta$ is negative, $K_{\rm zz}$ decreases with increasing pressure. In particular, within the same pressure range, the values of $K_{\rm zz}$ exhibit symmetry. This symmetry allows for clear differentiation of energy changes in the planet resulting from positive or negative values of $\zeta$.

\begin{figure}
    \centering
    \includegraphics[width=1.0\linewidth]{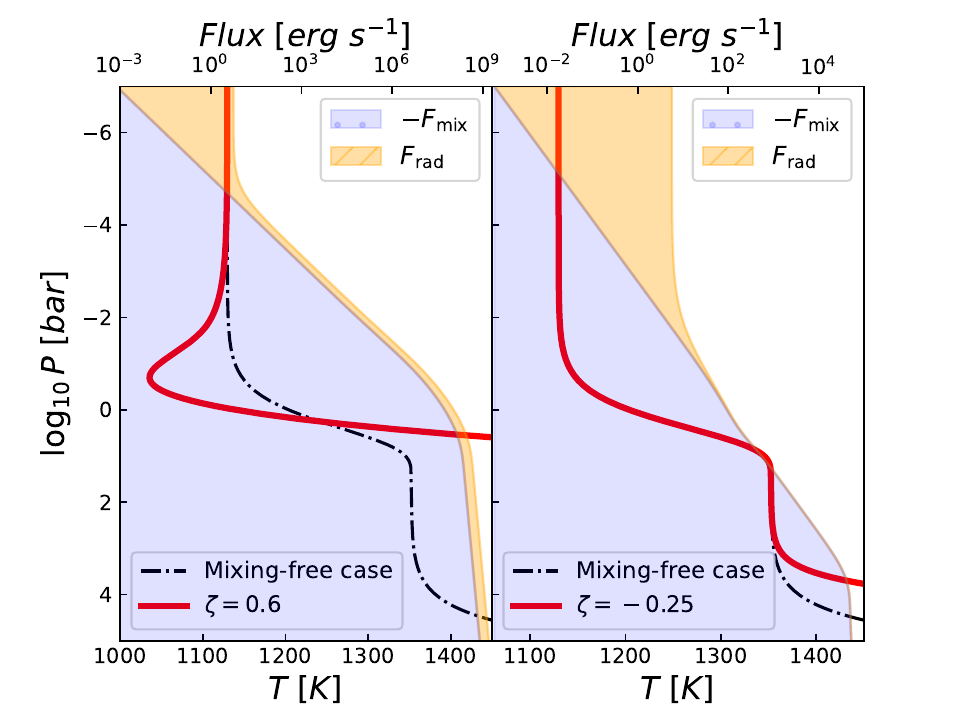}
    \caption{The division of mixing and radiative fluxes in the thermal structure models. The light purple area represents $\left|F_{\rm mix}\right|$, and the orange area represents $F_{\rm rad}$. According to the RME, $F_{\rm rad} = H/ 4\pi +  \left| F_{\rm mix} \right|$. The black dash-dot line represents the mixing-free scenario from Figure \ref{fig3}. The red line in the left panel shows $\zeta = 0.6$, while the right panel shows $\zeta = -0.25$.
   }
    \label{fig4_energy_flux}
\end{figure}

In the photosphere, dynamic winds create strong diffusion that weakens with altitude. A constant $K_{\rm zz} = 10^6 \ \text{cm}^2 \ \text{s}^{-1}$ heats the lower atmosphere, but this effect lessens as $\left|\zeta\right|$ increases. The interaction between the mixing flux ($F_{\rm mix}$) and the radiative flux ($F_{\rm rad}$) drives this behavior, as shown in the right panel of Figure \ref{fig4_energy_flux}. In the upper atmosphere, $F_{\rm mix}$ rises minimally, making $F_{\rm rad}$ dominant and stable. Consequently, the structures in both scenarios overlap. 
There is a critical point where the radiative flux  $\left|d F_{\rm rad}/dP \right|$ reach the maximum, marking the onset of heating in the deep atmosphere. At this point, the rate of change in $\left|F_{\rm mix}\right|$ starts to slow down. As $\left|\zeta\right|$ increases, this critical point moves deeper into the atmosphere, thereby reducing the extent of heating.
When $\left|\zeta\right| \ge 0.5$, the temperature-pressure profiles (green  and red dotted line) align with the mixing-free case.

In contrast, when vertical mixing is driven by shear interactions within convective zones, $\zeta>0$ \citep{2010ApJ...Youdin}. 
The detailed dynamical process: shear interaction drives vertical mixing by generating turbulence between fluid layers with different velocities, facilitating the exchange of heat, momentum, and substances across these layers.
Stronger diffusion occurs at greater depths, resulting in a more significant heating compared to a constant $K_{\rm zz}$. Higher $\zeta$ values can easily induce temperature inversions, as shown in Figure \ref{fig3}.
In the upper atmosphere, $\left|F_{\rm mix}\right|$ increases with pressure and is significantly higher with positive $\zeta$, compared to scenarios with negative $\zeta$, as illustrated in the left panel of Figure \ref{fig4_energy_flux}. This leads to substantial cooling due to increased $F_{\rm rad}$. 
In the deeper atmosphere, the initial decrease in $\left|dF_{\rm mix}/dP\right|$ in the shallower layers causes significant heating the atmosphere.  Consequently, alongside specific dynamic conditions, the mechanism of vertical mixing in these contexts promotes the emergence of temperature inversions within this parameter space.

In contrast to the effects of increasing versus decreasing $K_{\rm zz} $, scenarios with increasing $K_{\rm zz} $ intensify the warming of the deeper atmosphere without significantly cooling the upper atmosphere. 
These findings underscore the importance of accurately determining $\zeta$ within the parameter space to model the impact of spatially varying $K_{\rm zz }$ effectively and predict atmospheric dynamics accurately.

\begin{figure}
     \centering
      \includegraphics[width=9cm]{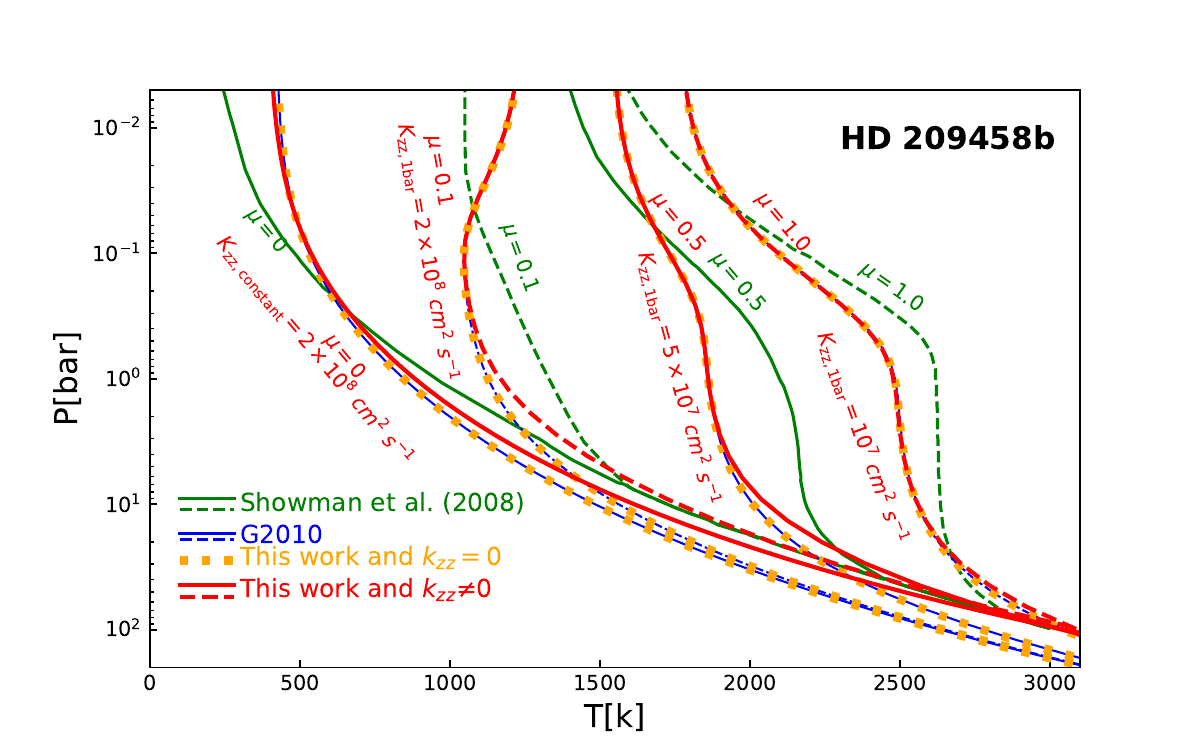}
     \caption{
  Temperature-pressure profiles  for various $\mu= \cos \theta$ values, compared with mixing effects and the calculations presented in \cite{2008ApJ..Showman} for HD~209458b. The orange lines are derived with $T_{\rm int}=500$ K. 
     The thick red lines illustrate the impact of vertical mixing.
     }
     \label{fig4}
\end{figure}


\subsection{Simulated Temperature Profiles for HD~209458b}
\label{sec3.3}

Prior research has concentrated on the mechanisms of vertical mixing induced by convective processes and turbulence within the lower atmospheric layers of Triton and Pluto.
In Triton, the mixing coefficient $K_{\rm zz}$ varies with depth \citep{1991Icar...Yelle}: It is about $10^6 \ \text{cm}^2/\text{s}$ below 8 km and $300 \ \text{cm}^2/\text{s}$ above that level. 
On Pluto, the one-dimensional heating transport model suggested that $K_{\rm zz}$ is approximately $10^{3} \ \text{cm}^2/\text{s}$ \citep{2021ApJ...Wan}, which agrees with the $K_{\rm zz}$ derived from 3D General Circulation Model (GCM) \citep{2020JGRE..12506120B...Bertrand} and various chemical models (refs in  \citealt{2021ApJ...Wan}).
Mixing in the upper atmosphere is less explored, although vertical mixing by gravity wave dissipation has been verified by measurements of the Galileo Atmospheric Structure Instrument \citep{2005Icar..Young}. 
Moreover, the strength of $K_{\rm zz}$ in the upper atmosphere is model-dependent. 
{Here, we test $K_{\rm zz}$ in the upper atmosphere of HD~209458b, assuming vertical mixing is driven by gravity waves or atmospheric circulation.

In the absence of vertical mixing, G\citeyear{2010A&A...Guillot} matched the models of \citet{2008ApJ...Fortney} and \citet{2008ApJ..Showman} by keeping $g = 980 \ cm \ s^{-2}$ and adjusting the opacities in the visible and thermal bands ($\kappa_{\rm v} = 4\times 10^{-3} \ g \ cm^{-2}$ and $\kappa_{\rm th} = 10^{-2} \ g \ cm^{-2}$). The discrepancies between G\citeyear{2010A&A...Guillot} and other models are of the same order as the differences among the models themselves. Vertical mixing may change these discrepancies. This section explores whether varying the intensity of mixing can better match the atmospheric structure of HD~209458b, as shown in Figure 1 of \cite{2008ApJ..Showman}.

To verify our simulation, we compared our simulation results (orange dotted line) with the analysis results (blue lines). 
These two lines overlap, but differ from the green line.
In Figure \ref{fig4}, the blue lines show the adjusted radiative profile to match the green lines with $T_{\rm irr} = 1469$ K on the night-side structure, using similar transparency settings as in Figure \ref{fig2}. 
For $T_{\rm irr} = 2000$ K at the other incident angle cosine, the settings match those in Figure \ref{fig1_modifies} (left). 
The intrinsic effective temperature of the planet is $T_{\rm int} = 500$ K. 
Despite smaller differences when $\mu < 0.3$, a structural gap remains. 

Vertical mixing can reduce the temperature difference.
Typically, the estimation of $K_{\rm zz}$ encompasses heat transport \citep{2017Natur.551..352Z}, chemical processes, and dynamic interactions. However, this study exclusively considers the energy transport due to vertical mixing. Consequently, the associated heat transport also changes, leading to an eddy Prandtl number that deviates from unity.
It should be noted that the Prandtl number assesses the relative significance of thermal conduction in comparison to momentum transport \citep{Zhang_2020}.
The energy transport of the vertical mixing will determine the value of $K_{\rm zz}$. 

For HD~209458b, a reasonable vertical diffusion profile is $K_{\rm z z} \propto \left({P}/{1 \mathrm{bar}}\right)^{-1/2} \mathrm{~cm}^2 \mathrm{~s}^{-1}$ in an isothermal atmosphere \citep{lindzen1981turbulence,Zhang_2020} for $\mu>0$. 
In a deeper atmosphere, the heating effect of vertical mixing increases with the cosine of the incident angle. Therefore, the mixing strength at the top atmosphere must decrease with this cosine to effectively overlap with the lower layer of HD~209458b. 
\cite{2019MNRAS...Menou} suggested that a reasonable vertical diffusion profile for HD~209458b is $K_{\rm zz} \simeq 10^9 \left(P/1 \ {\rm bar} \right)^{-2} \ \rm cm^2 \ s^{-1}$. 
In line with this,  the strength of  $K_{\rm zz, 1bar} \simeq 10^{7}-10^{8.3} \ \rm cm^2 \ s^{-1}$ at 1 bar in our model is less than the vertical diffusion noted in the introduction. 
Moreover,  a 
suitable constant $K_{\rm zz}$ heats the lower atmosphere. 
When $\mu = 0$,  
a constant  $K_{\rm zz} = 2 \times 10^8 \ \rm cm^2 \ s^{-1}$ is sufficient to cover the lower layer.
However, these vertical mixing strengths in this section are insufficient to provide the necessary energy for cooling the upper atmosphere, which may dissipate due to dynamic effects  or the chemical components. 
In this section, the lower atmosphere temperature mainly determines the mixing strength at  1 bar. Significant cooling in the upper atmosphere, shown in Figure~\ref{fig1_modified_02}, requires stronger mixing at higher altitudes. However, as seen in Figure~\ref{fig4}, the warming of the lower atmosphere is less pronounced compared to the strong mixing conditions of Figure~\ref{fig1_modified_02}. Consequently, the parameter configuration illustrated in Figure~\ref{fig4} does not facilitate a reduction in the temperature of the upper atmosphere.


\section{conclusion}\label{sec_conclusion}


We incorporate the vertical mixing into a radiative equilibrium atmosphere to establish a radiative-mixing equilibrium atmosphere.
In this work, we examine whether the vertical mixing can enhance the development of  a temperature inversion in the planet atmosphere. 
It also estimates the vertical mixing strength based on the atmospheric structure of HD~209458b, as suggested by \cite{2008ApJ..Showman}.

Vertical mixing significantly influences temperature profiles, where $\mu$, $\gamma$, and $K_{\rm zz}$ collectively dictate its effects. In conditions where $\gamma > 1$ and $\mu < 0.5$, temperature inversions naturally occur in the absence of mixing, but these are mitigated by robust, continuous mixing. 
Conversely, when 
$\mu \ge 0.5$, temperature profiles generally exhibit a steady increase 
with increasing pressure for the cases without mixing. 
Vertical mixing raises temperatures at greater depths, reducing temperature inversions as $\mu < 0.5$ and promoting a gradual temperature increase as $\mu \ge 0.5$. It also creates pseudo-adiabatic regions, shifting the radiative-convective boundary (RCB) inward. 
A minor temperature inversion is induced by vertical mixing, as the temperature profiles with increasing pressure are established with reduced thermal opacity.
Spatial variations in $K_{\rm zz}$ can dynamically intensify temperature inversions or heat deeper layers in certain circumstances. 
For HD 209458b, a diffusion profile $K_{\rm zz} \propto (P/1\ \text{bar})^{-1/2} \ \text{cm}^2 \ \text{s}^{-1}$ is applied, indicating a mixing strength of approximately $10^{7}-10^{8.3} \ \text{cm}^2 \ \text{s}^{-1}$  at 1 bar, consistent with its deep atmospheric structure. This study emphasizes the critical role of vertical mixing in shaping temperature profiles, highlighting a decrease in mixing strength  compared to previous findings.

The current RME model has shown promise in practical applications, but it has some limitations and prospects. 
These limitations arise primarily from the lack of consideration of the actual convective zone and chemical equilibrium in non-grey atmospheres.
Firstly, by incorporating the  actual convective zone, we can achieve a more precise evaluation of the temperature gradients of the planet \citep{2012ApJ...Robinson,2015A&A...Parmentier,2023ApJ...Zhang}. Secondly, chemical elements play a crucial role in planetary atmospheres as they not only influence the atmospheric climate \citep{2010.book..Pierrehumbert}, but also modify the structure of the atmosphere through interactions with clouds \citep{2017MNRAS...Hubeny}. 
Chemical elements are very sensitive to temperature changes, with even small fluctuations having a large impact. We have shown that vertical mixing significantly affects temperature, suggesting that chemical elements will shift noticeably.
In addition, the RME under general conditions could affect a planet's runaway greenhouse effect \citep{2023Natur.Selsis} by altering the levels of water vapor and the allocation of carbon and hydrogen within the magma ocean.

Our atmosphere model is based on the Gensis model \citep{2017MNRAS...GENESIS,Piette...phdthesis}, which can effectively incorporate considerations of chemical equilibrium in multi-band frequency. 
These enhancements will provide us with better insight into exoplanet habitability, climate, transition radius \citep{2008ApJS..Hansen}, observational characteristics (transmission spectrum,  \citealt{2022AJ....Zhou}), while having significant implications for explaining planetary mass and water content within the TRAPPIST-1 system. 
Furthermore, both isotropic and nonisotropic scattering \citep{2014ApJS..215...Heng} can also alter the temperature-pressure profile of the atmosphere.
Furthermore, RME can shift the position of the photosphere, significantly influencing the mass loss of the envelope through photoevaporation \citep{2023ApJ...Modirrousta}.

\section{acknowledgments}
     We thank the anonymous referee, Xi Zhang, and V. Parmentier for their valuable suggestions that greatly improved this work.
     This work has been supported by the National SKA Program of China (grant No. 2022SKA0120101) and National Key R \& D Program of China (No. 2020YFC2201200) and the science research grants from the China Manned Space Project (No. CMS-CSST-2021-B09, CMS-CSST-2021-B12, and CMS-CSST-2021-A10), Guangdong Basic and Applied Basic Research Foundation (grant 2023A1515110805), and the grants from the opening fund of State Key Laboratory of Lunar and Planetary Sciences (Macau University of Science and Technology, Macau FDCT Grant No. SKL-LPS(MUST)-2021-2023). C.Y. has been supported by the National Natural Science Foundation of China (grants 11521303, 11733010, 11873103, and 12373071).   
     B.M. has been supported by the National Natural Science Foundation of China (grants:12073092).

\bibliography{sample631}{}
\bibliographystyle{aasjournal}

\end{document}